\newif\ifSUBMIT
\newif\ifCOMMENTS

\let\ifSUBMIT\iffalse
\let\ifCOMMENTS\iftrue

\ifSUBMIT
    \newcommand*\DocClsOpts{preprint}
\else
    \newcommand*\DocClsOpts{reprint}
\fi

\documentclass[
    aps,
    pre,
    \DocClsOpts,
    amsmath,
    amssymb,
    superscriptaddress,
    floatfix,
]{revtex4-1}

\usepackage[utf8]{inputenc} 

\usepackage{etoolbox} 
\usepackage{xparse}

\usepackage{amsmath}
\usepackage{amssymb}
\usepackage{mathtools}
\usepackage{bm}
\usepackage{bbold}
\usepackage{physics} 
\usepackage{siunitx} 

\usepackage{csquotes} 
\usepackage[inline]{enumitem}
\usepackage{graphicx}
\usepackage{xcolor} 

\AtBeginDocument{\usepackage{booktabs}} 

\usepackage{todonotes}

\usepackage{acronym}
\usepackage{hyperref}

\definecolor{edits}{RGB}{200,0,0}
\definecolor{strike}{RGB}{130,0,0}
\ifSUBMIT
    \ifCOMMENTS
        \usepackage[normalem]{ulem}
        
        \NewDocumentCommand\STRIKE{+m}{{\color{strike}\sout{#1}}}
    \else
        
        \NewDocumentCommand\STRIKE{+m}{}
    \fi
\else
    \usepackage[normalem]{ulem}
    
    \NewDocumentCommand\STRIKE{+m}{{\color{strike}\sout{#1}}}
\fi

\NewDocumentCommand\eg{}{e.\,g.}
\NewDocumentCommand\ie{}{i.\,e.}
\NewDocumentCommand\cf{}{cf.}

\NewDocumentCommand\MathPeriod{}{\,\text{.}}
\NewDocumentCommand\MathComma{}{\,\text{,}}

\NewDocumentCommand\ee{}{\mathrm{e}}

\newlength\figurewide
\AfterEndPreamble
{
    \ifSUBMIT
        \setlength\figurewide{0.6\columnwidth}
    \else
        \setlength\figurewide{\columnwidth}
    \fi
}

\NewDocumentCommand\LD{}{\ensuremath{\mathcal{L}}}
\NewDocumentCommand\Ws{}{\ensuremath{\mathcal{W}_\mathrm{s}}}
\NewDocumentCommand\Wu{}{\ensuremath{\mathcal{W}_\mathrm{u}}}

\NewDocumentCommand\xDS{}{\ensuremath{x^\mathrm{DS}}}

\NewDocumentCommand\Nr{}{\ensuremath{N_\mathrm{R}}}
\NewDocumentCommand\nmc{}{\ensuremath{n_\mathrm{mc}}}
\NewDocumentCommand\kAvrg{}{\ensuremath{\bar{k}}}
\NewDocumentCommand\kEnse{}{\ensuremath{k_\mathrm{e}}}
\NewDocumentCommand\kEnseAvrg{}{\ensuremath{\bar{k}_\mathrm{e}}}
\NewDocumentCommand\kMani{}{\ensuremath{k_\mathrm{m}}}
\NewDocumentCommand\kManiAvrg{}{\ensuremath{\bar{k}_\mathrm{m}}}
\NewDocumentCommand\kFloq{}{\ensuremath{k_\mathrm{F}}}
\NewDocumentCommand\kFloqLoc{}{\ensuremath{k_\mathrm{F}^\mathrm{loc}}}

\NewDocumentCommand\RR{mom}{\ensuremath{\mathrm{#1} \IfValueT{#2}{\to \mathrm{#2}} \to \mathrm{#3}}}

\begin{document}

\title{Dynamics and decay rates of a time-dependent two-saddle system}

\author{Johannes Reiff}
\author{Matthias Feldmaier}
\author{Jörg Main}
\affiliation{
    Institut für Theoretische Physik I,
    Universität Stuttgart,
    70550 Stuttgart, Germany
}

\author{Rigoberto Hernandez}
\email[Correspondence to: ]{r.hernandez@jhu.edu}
\affiliation{%
    Department of Chemistry,
    Johns Hopkins University,
    Baltimore, Maryland 21218, United States
}
\affiliation{%
    Departments of Chemical \& Biomolecular Engineering,
    and Materials Science and Engineering,
    Johns Hopkins University,
    Baltimore, Maryland 21218, United States
}

\date{\today}

\begin{abstract}
    \label{sec:abstract}

    The framework of transition state theory (TST) provides a powerful way for
    analyzing the dynamics of physical and chemical reactions. While TST has
    already been successfully used to obtain reaction rates for systems with a
    single time-dependent saddle point, multiple driven saddles have proven
    challenging because of their fractal-like phase space structure. This paper
    presents the construction of an approximately recrossing-free dividing
    surface based on the normally hyperbolic invariant manifold in a
    time-dependent two-saddle model system. Based on this, multiple methods for
    obtaining instantaneous (time-resolved) decay rates of the underlying
    activated complex are presented and their results discussed.
\end{abstract}

\maketitle


\acrodef{BCM}{binary contraction method}
\acrodef{DoF}{degree of freedom}
\acrodefplural{DoF}{degrees of freedom}
\acrodef{DS}{dividing surface}
\acrodef{LD}{Lagrangian descriptor}
\acrodef{LMA}{local manifold analysis}
\acrodef{NHIM}{normally hyperbolic invariant manifold}
\acrodef{PES}{potential energy surface}
\acrodef{TS}{transition state}
\acrodef{TST}{transition state theory}


\section{Introduction}
\label{sec:intro}

One of the central aims in the field of chemical reaction dynamics
is the accurate determination of reaction rates.
This is not just an abstract problem of academic (or basic) concern%
~\cite{truh83, marc92, truh96, KomatsuzakiBerry01a},
but also a practical problem
with many potential applications in complex
reactions~\cite{dougherty06, green11, henkelman14, Ezra2015, hoffmann17}.
The possibility of optimizing reaction rates by external driving
could perhaps take these applications further
in offering improvements to throughput and efficiency.

Multi-barrier reactions were considered early~\cite{mill68, mill79b}
in the context of quantum mechanical tunneling through barriers
at constant energy.
In this \emph{M-problem}, the periodic orbit between the barriers
gives rise to the possibility of an infinite number of
returns to the turning point from which tunneling can proceed.
The return times are usually not commensurate with the period,
either because of coupling to
other degrees of freedom---such as from the bath---or
because of variations in the potential.
In such cases, the
coherence in the returns is altered, changing the nature of the dynamics
in ways that we address in this work.

Problems involving fluctuating~\cite{DG92, Mad92}
or oscillating~\cite{Lehmann00a, Lehmann00b, Lehmann03} barriers
have also received significant attention leading to, for example,
the identification of the phenomenon of resonant activation~\cite{VDB93, Hang94}.
While the approaches originally focused
on the overdamped regime~\cite{DG92, Lehmann00a},
underdamped systems were later examined~\cite{hern01d, Lehmann03}.
For example, mean first passage times have been employed
to calculate (diffusion) rates in spatially periodic multi-barrier potentials.
Therein, various static~\cite{hern02b}
as well as stochastically driven~\cite{hern02c, hern04c} cases
have been characterized primarily through numerical methods.

In the current context, the main challenge in a multi-saddle system
comes from the unpredictability of states in the intermediate basin.
A reactant entering this region may leave
either as a reactant or product depending on the exact
initial conditions~\cite{DeVogelaere1955, mill76b, pollak80, Carpenter2005a}.
Historically, this challenge has been approached by
categorizing reactions into two classes~\cite{mill76b, pech76, pech81}:
\emph{Direct} reactions exhibit a single \ac{TS}.
\emph{Complex} reactions, on the other hand,
have two clearly separated \acp{TS}.
The potential well between those barriers
is assumed to be sufficiently deep
that it gives rise to a long-lived collision complex.
Trajectories passing through one \ac{TS} enter this collision complex and
hence cannot be correlated to trajectories passing through the other \ac{TS}.
In reality, however, a reaction cannot always be uniquely classified.
These concerns were addressed in a unified theory by Miller~\cite{mill76b},
and later refined by Pollak and Pechukas~\cite{pollak79}
so as to address shallow potential wells.
While an important advancement,
this theory still treats the saddle's interactions statistically,
thereby neglecting dynamical effects like resonances.
Moreover, they considered
multi-step reactions in which the positions and heights of the barriers
are time-independent.
Last, there are numerous publications on valley-ridge inflection points,
which are typically described by a normal \ac{TS} followed by
a \emph{shared} one~\cite{Carpenter2005a, Rehbein2011, Collins2013}.

Craven and Hernandez~\cite{hern16d} recently examined
a four-saddle model of ketene isomerization
influenced by a time-dependent external field.
They encountered complicated phase space structures similar to
those in systems with closed reactant or product basins~\cite{hern17e}.
As a result, their analysis was limited to local \acp{DS}
and no reaction rates were calculated.
Moreover, successfully calculating instantaneous rates
based on a globally recrossing-free \ac{DS}
attached to the \ac{NHIM} of a time-dependent multi-saddle system
has---to our knowledge---not yet been reported.

In this paper, we address
the challenge of determining the instantaneous \ac{TS} decay rate
for systems that not only feature multiple barriers along the reaction path,
but that are also time-dependent.
Specifically,
we investigate an open, time-dependent two-saddle model with one \ac{DoF}
as introduced in Sec.~\ref{sec:materials/model}.
The theoretical framework
along with the numerical methods used throughout the paper
are described in Secs.~\ref{sec:materials/ds_nhim}
through~\ref{sec:materials/rates}.
We then discuss the system's phase space structure
under the influence of periodic driving at different frequencies
in Sec.~\ref{sec:ps}.
By leveraging unstable trajectories bound between the saddles,
we can construct an almost globally recrossing-free \ac{DS}
associated with the so-called geometric cross
in Sec.~\ref{sec:cross}.
The \ac{DS} is used
in Sec.~\ref{sec:rates}
to calculate instantaneous decay rates and averaged rate constants
for the underlying activated complex using different methods.
Thus we report a
detailed analysis of the nontrivial phase space structure
of the chemical M-problem
and the calculation of its associated decay rates.


\section{Methods and materials}
\label{sec:materials}


\subsection{Time-dependent two-saddle model system}
\label{sec:materials/model}

In this paper, we investigate
the properties of multi-barrier systems by considering
a 1-\ac{DoF} model potential featuring two Gaussian barriers whose
saddle points are centered at $x = \num{+-1}$.
Initially, both barriers are placed at the same level.
As we are interested in considering the time-dependent case, however,
we drive the barrier's heights $B_\varphi(t)$ sinusoidally in opposite phases.
That is, we use the same amplitude and frequency $\omega$ for both saddles,
but opposite initial phases $\varphi \in \qty{0, \pi}$.
This leads to the potential
\begin{subequations}
    \label{eq:materials/model}
    \begin{align}
        V(x, t) &= B_0(t) \ee^{-(x + 1)^2} + B_\pi(t) \ee^{-(x - 1)^2} \\
        \qq*{with}
        B_\varphi(t) &= \frac{7}{4} + \frac{1}{4} \sin(\omega t + \varphi)
        \MathPeriod
    \end{align}
\end{subequations}
The oscillation frequency $\omega$ is a free parameter
that can be varied relative to the other natural time scales of the system
at a given fixed total energy.
At an arbitrary time, one of the barriers will be larger than the other.
For example, when the second barrier is larger,
the potential takes a shape such as that shown at the top
of Fig.~\ref{fig:static_manifolds}(b).
Throughout this paper,
dimensionless units are used to explore the range of phenomena that can arise
from varying the relative timescales of the system and the driving.


\subsection{Dividing surfaces and the NHIM}
\label{sec:materials/ds_nhim}

\begin{figure}
    \includegraphics[width=\figurewide]{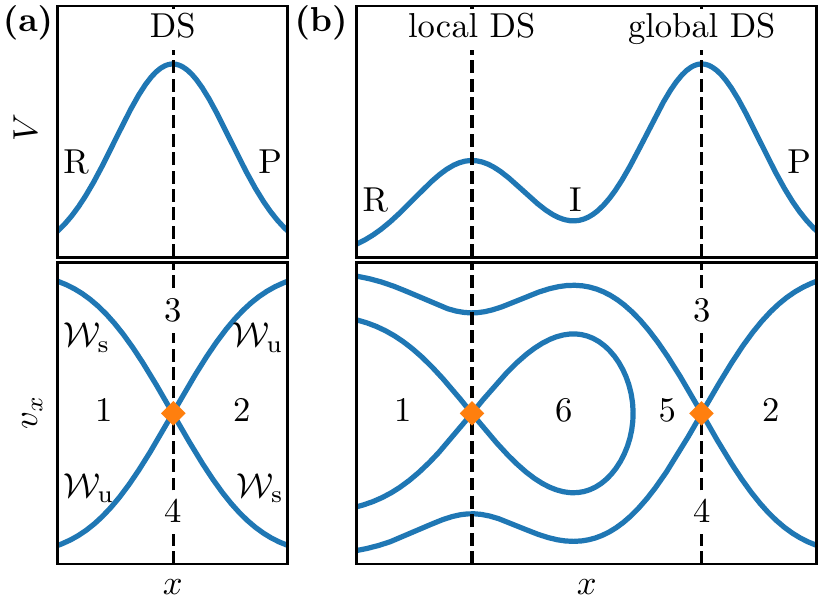}
    \caption{%
        Typical structures of static potentials $V(x)$
        and their corresponding phase spaces $x$--$v_x$
        with one and two barriers in panels (a) and (b), respectively.
        The potential barriers separate reactant (R) and product (P) states.
        The two-barrier case features an additional intermediate (I) state
        in between.
        The maxima are associated with a hyperbolic fixed point (diamonds)
        and a dividing surface (indicated by dashed vertical lines) each.
        The corresponding manifolds divide the phase space into
        four distinct, numbered regions in panel (a)
        and six regions in panel (b).
        See Secs.~\ref{sec:materials} and~\ref{sec:ps/limits}
        for details.}
    \label{fig:static_manifolds}
\end{figure}

Reactants and products on the canonical \ac{PES}
are usually separated by a \ac{DS} often associated with a rank-1 saddle,
such as that shown in Fig.~\ref{fig:static_manifolds}(a),
whose unstable direction identifies the reaction coordinate.
The maximum energy configuration along
the corresponding one-dimensional minimum energy
path~\cite{Glasstone1941, Fukui1970, Fukui1976}
$x$ is called the \ac{TS}~\cite{eyring35, wigner37, pech81, truh96}.
In \ac{TST}, the local decay rate is obtained from the flux through the \ac{DS}.

A \ac{DS} is locally recrossing-free if no particle pierces it more than once
before leaving some pre-determined interaction region around the saddle.
In this case, the decay rates (to exit the interaction region)
as determined by the \ac{DS} are locally exact.
We refer to a \ac{DS} as being
globally exact or recrossing-free
if the above is true independent of the choice of the interaction region
as long as said regions do not overlap
with the stable reactant or product regions~\cite{hern19a}.
Using this definition avoids inherent recrossings
caused by reflections in closed reactant or product basins~\cite{hern17e}.
In this paper,
we only address transitions over barriers in series,
and we do not address the parallel case in which a reaction could
access more than one distant barrier.
The scope of the definitions of globally exact and recrossing-free
is therefore limited accordingly.

Every saddle point of a $d$-\ac{DoF} potential
is associated with $(2 d - 1)$-dimensional
stable and unstable manifolds \Ws\ and \Wu\ in phase
space---\cf\ Fig.~\ref{fig:static_manifolds}(a).
Their $(2 d - 2)$-dimensional intersection is called the
\ac{NHIM}~\cite{Lichtenberg82, hern93b, hern93c, Ott2002a, wiggins2013normally}
and describes the unstable subspace of particles trapped on the saddle.
It is noted with a diamond in Fig.~\ref{fig:static_manifolds}(a).
Depending on the convention used,
the time-evolution of the \ac{NHIM} or of a single point on the \ac{NHIM}
forms the \ac{TS} trajectory~\cite{hern19a}.
A ($2 d - 1$)-dimensional \ac{DS} can be constructed
by attaching it to the \ac{NHIM},
which works even for time-dependent driving.
This particular \ac{DS} is locally recrossing-free
as long as particles do not cross far away from the
\ac{NHIM}~\cite{KomatsuzakiBerry99, komatsuzaki06b, komatsuzaki06a, Li09}.
However, it does not have to be globally recrossing-free if, \eg,
the system features multiple barriers or valleys with sharp turns.

In \ac{TST}, one typically uses the saddle
point as the \ac{TS} itself, but the variational principle
clearly suggests that the \ac{DS} can be moved away from it.
In practice, this has led to applications in which the \ac{DS}
is associated with the saddle point.
However, not only is this strong association not necessary,
it is even possible that the \ac{NHIM} (which anchors the \ac{DS})
can be disconnected.
Indeed, we find in the current problem that the optimal \ac{NHIM}
consists of multiple, disjoint sets.
As seen in Sec.~\ref{sec:ps},
such structure can emerge
from the interaction of multiple saddles.
Those parts of the \ac{NHIM} not associated with a saddle
can nevertheless be used to construct a \ac{DS} as before,
and may feature fewer global recrossings compared to
a \ac{DS} associated uniquely with a single saddle point.
The question of whether such a \ac{DS} with fewer or even no recrossings
can be found in a driven multi-saddle system
has led to the central results of this work.


\subsection{Revealing geometric structures}
\label{sec:materials/geometry}

The geometric structure of the phase space can be revealed
using the \ac{LD}~\cite{Mancho2010, Mancho2013, hern15e, hern16a, hern16d}
defined by
\begin{equation}
    \label{eq:materials/ld}
    \LD(x_0, v_0, t_0) = \int_{t_0 - \tau}^{t_0 + \tau} \dd{t} \norm{v(t)}
\end{equation}
for a given initial position $x_0$, velocity $v_0$, and time $t_0$.
It measures the arc length of a trajectory $x(t)$
in the time interval $t_0 - \tau \le t \le t_0 + \tau$.
A local minimum in the \ac{LD} arises
when the particle covers the minimum distance
in the interval $t_0 - \tau \le t \le t_0 + \tau$.
It consequently remains longer in the interaction region
when integrating forwards or backwards in time,
and thus provides a signature for the presence of a stable or unstable manifold,
respectively.

The \ac{LD} has the advantage that it is conceptually very simple
and that it can be applied to practically any system.
This makes it suitable for a first visual inspection.
As discussed in Ref.~\cite{hern18g}, however,
it features a nontrivial internal structure.
Numerically determining the exact position of stable and unstable manifolds
is therefore difficult.

A numerically simpler scheme is based on
the concept of reactive (and nonreactive) regions
as described in Refs.~\cite{hern18g, hern19a}.
It discriminates initial conditions
by first defining an interaction region in position space
that encompasses the relevant dynamics.
Particles are then propagated forwards and backwards in time
until they leave said interaction region.
In both directions of time,
a particle can end up as either reactant (R) or product (P).
This leads to four possible classifications for a
given initial condition
as shown in Fig.~\ref{fig:static_manifolds}(a)
for each of the four regions:
\begin{enumerate*}[label=\arabic*.]
    \item nonreactive reactants \RR{R}{R},
    \item nonreactive products \RR{P}{P},
    \item reactive reactants \RR{R}{P}, and
    \item reactive products \RR{P}{R}.
\end{enumerate*}
Similar concepts have been introduced in, \eg, Ref.~\cite{hern16d}.

Such a classification for the initial conditions has to be extended
to include the consequences of a local minimum between the saddles of the
reacting system of Eq.~\eqref{eq:materials/model}.
A low-energy particle trapped near this local minimum
[\cf\ Fig.~\ref{fig:static_manifolds}(b)]
would lead to diverging computation times
because it may never leave the interaction region.
To solve this problem, an additional termination condition is introduced,
whereby any particle that crosses the potential minimum
a specified number of times \nmc\
is classified as an intermediate (I) particle.
Consequently, up to nine different regions in phase space can be distinguished
for any given value of \nmc.

Using the concept of reactive regions,
stable and unstable manifolds can be revealed
as borders between adjacent regions.
Their closure's intersections form the \ac{NHIM}.
The algorithm used to calculate points of the \ac{NHIM}
is based on the \ac{BCM} introduced in Ref.~\cite{hern18g}.
It starts by defining a quadrangle
with one corner in each of the phase space regions
surrounding the manifold intersection.
The quadrangle is then contracted
by successively determining the region corresponding to an edge's midpoint
and moving the appropriate adjacent corner there.
The \ac{BCM} is therefore effectively composed of
four intertwined classical bisection algorithms.
To reliably identify the initial corners,
we modify the \ac{BCM} slightly as detailed in Appendix~\ref{sec:bcm}.


\subsection{Calculating decay rates}
\label{sec:materials/rates}

The existence of a \ac{NHIM} of codimension 2 and its role in determining
the chemical reaction rate brings an additional concern.
Namely, what is the degree of instability of the \ac{TS} as determined
by the decay of trajectories that start in the proximity of the \ac{NHIM}?
In a time-dependent---\eg\ driven---environment,
this instantaneous decay will be time-dependent as well.
Nevertheless, it can be assigned a single characteristic decay rate constant
when the time-dependence is periodic
by taking the average over the period~\cite{Lehmann00a, hern19e}.

In this paper, we implement three different
methods, summarized in Appendix~\ref{sec:ratemethods},
for calculating decay rates:
\begin{enumerate*}[label=\emph{(\roman*)}]
\item
    The ensemble method~\cite{Lehmann00a, hern19e}
    yields instantaneous (time-resolved) rates
    by propagating a large number of particles.
    It is computationally expensive but conceptually simple.
\item
    The \ac{LMA}~\cite{hern19e, hern20m}
    accelerates the computation of instantaneous rates
    by leveraging the linearized dynamics near the \ac{NHIM}.
\item
    If only average rate constants are desired,
    the Floquet rate method~\cite{hern14f, hern19e} can be used instead
    while requiring even less computational resources.
\end{enumerate*}

In the cases resolving the dynamics of
the two-barrier problem of Eq.~\eqref{eq:materials/model},
all three generally converge within reasonable time.
However, they each involve different assumptions which might
have led to different results, and which can provide
complementary interpretations about the underlying dynamics.
As shown in the results sections, all three lead to
decay rates in excellent agreement.
The repetition thus also serves to provide assurance in the
reported values.


\section{Geometric structure of the two-saddle system}
\label{sec:ps}

The phase space structure of
the model system introduced in Sec.~\ref{sec:materials/model}
is highly dependent on the driving frequency $\omega$.
In the following we will give
a qualitative overview of the behavior the system can exhibit.


\subsection{Limiting cases}
\label{sec:ps/limits}

A static two-saddle system
akin to Eq.~\eqref{eq:materials/model} with $\omega = 0$
exhibits the phase space structure shown
in Fig.~\ref{fig:static_manifolds}(b).
The saddle tops are associated with hyperbolic fixed points
whose stable and unstable manifolds each form a cross.
If the first saddle is smaller than the second one,
two of its manifolds constitute a homoclinic orbit.
In this case, the phase space is composed of six regions, namely
\begin{enumerate*}[label=\arabic*.]
    \item nonreactive reactants \RR{R}{R},
    \item nonreactive products \RR{P}{P},
    \item reactive reactants \RR{R}{P},
    \item reactive products \RR{P}{R},
    \item particles that react over the first saddle
        but get reflected at the second \RR{R}[I]{R}, and
    \item intermediate particles that are trapped between the saddles \RR{I}{I}.
\end{enumerate*}

Likewise, if the driving frequency is sufficiently large ($\omega \to \infty$),
the particle will effectively see an average static potential
in which it must cross two similar static barriers of equal height.
As the energy is conserved,
once the particle crosses the first barrier,
it necessarily crosses the second barrier.
This results in a phase space structure similar to that for the static case of
Fig.~\ref{fig:static_manifolds}(b)---although with
heteroclinic orbits connecting the hyperbolic fixed points.

In such (effectively) static cases,
it is straightforward to define a global recrossing-free \ac{DS}.
In a 1-\ac{DoF} constant energy system,
if (and only if) a particle crosses the highest saddle,
it has demonstrated to have enough energy to react over all saddles.
Since the largest barrier therefore unambiguously determines
whether a particle reacts or not,
its associated local \ac{DS} becomes the global (recrossing-free) \ac{DS}.

While this holds true for static systems,
dynamically driven systems may exhibit much richer dynamics.
For example, in the case of an alternating pair of dominant barriers,
as in the model of Eq.~\eqref{eq:materials/model},
the naive \ac{DS} jumps discontinuously
from one side to the other twice per period.
As a result there exist reactive trajectories that never cross the \ac{DS}.
Instead, the latter jumps over the former
leading to an inconsistent description of the reaction.
Addressing this issue by defining particles between the local \acp{DS}
as reacting the moment the dominant saddle changes would lead to more problems,
\eg, unphysical Dirac delta peaks in the reaction rate.
To solve this issue, the system has to be treated as a whole.


\subsection{Intermediate driving frequency}
\label{sec:ps/medium}

\begin{figure}
    \includegraphics[width=\figurewide]{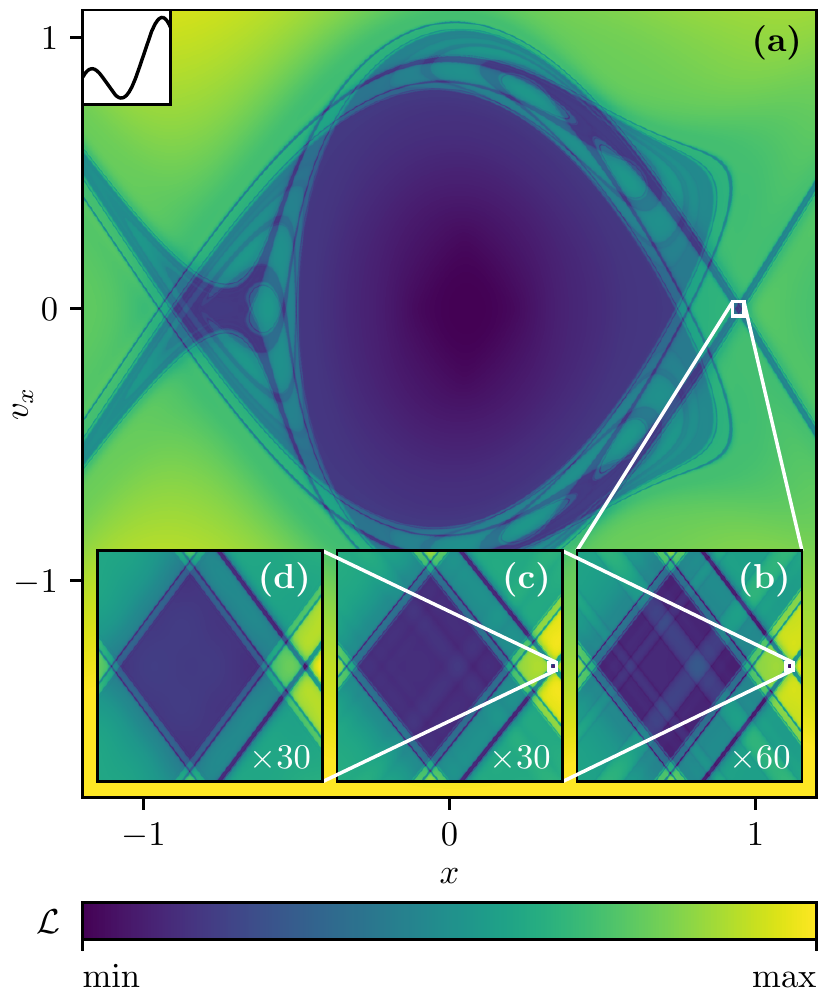}
    \caption{%
        Phase space structure for the time-dependent potential
        in Eq.~\eqref{eq:materials/model}
        with $\omega = \pi$ at $t_0 = 3 / 2$
        as revealed by the \acl{LD} \LD\ given in
        Eq.~\eqref{eq:materials/ld} with $\tau = 16$.
        Although the two geometric crosses
        seen in Fig.~\ref{fig:static_manifolds}(b)
        are still present in panel (a),
        they now exhibit a complicated substructure
        involving a vast number of homoclinic and heteroclinic points
        as well as homoclinic and heteroclinic orbits.
        The progressively zoomed cutouts
        shown in subpanels (b) through (d) exhibit self-recurring structures.
        Labels in the bottom right corners
        indicate the corresponding enlargement from the previous zoom level.
        The potential at time $t_0$
        is indicated in the top left inset of panel (a).}
    \label{fig:fractal_ld}
\end{figure}

The limiting cases discussed so far result in effectively static systems.
Since we are interested in novel and nontrivial behavior, however,
we will now turn to intermediate driving frequencies.
These can exhibit varying degrees of complexity
as a function of the driving frequency $\omega$.
An example of such non-trivial behavior with a highly complex phase space
is shown in Fig.~\ref{fig:fractal_ld}.
We use the \ac{LD} defined in Sec.~\ref{sec:materials/geometry}
for visualization
since it is very general and requires little knowledge about the system.

The geometric structure
was obtained for the driven potential $V(x, t)$
of Eq.~\eqref{eq:materials/model}
at an intermediate driving frequency $\omega = \pi$.
The general shape of the boundaries
separating reactive and nonreactive regions
(\cf\ Fig.~\ref{fig:static_manifolds}(b))
is still vaguely visible.
However, the precise position
of the crossing points between the stable and unstable manifolds
can no longer be determined.
This family of crossing points
together with the associated stable and unstable manifolds
within their vicinity
appears as a cross that has arisen from all of these geometric considerations.
For simplicity, we define it as a \emph{geometric cross} throughout this work.
Note that this term is not meant to be a precise mathematical
  structure but rather an illustrative concept for describing the
  complex phase space of the system under study.

The fractal-like geometric crosses
seen in the series of Figs.~\ref{fig:fractal_ld}(a)--(d)
for a finite $\tau$
suggests a fractal structure at all scales
for $\tau \to \infty$.
This structure emerges from particles trapped between the two saddles.
For example,
a reactant can enter the intermediate region over the left saddle,
be reflected multiple times at both saddles,
and finally leave the interaction region over the right saddle as a product.
The number of reflections
is in this case highly dependent on the particular initial conditions
as a result of the system being chaotic.
In turn, this leads to a discontinuity in the \ac{LD},
and eventually to the self-recurring patterns of a fractal structure.

Figure~\ref{fig:fractal_ld} also supports
the observation of geometric structures at a given time
that are thinner near the dominant saddle compared to the lower energy saddle.
In Fig.~\ref{fig:fractal_ld}(a), when the right barrier is dominant,
particles initialized near the higher saddle
start with higher potential energy
and therefore have a lower chance of being reflected.
As a result, fewer of these particles linger in the interaction region and
the phase space structures are thinner.
While this may be an interesting observation,
the geometric cross near the dominant barrier is still highly fractal.
There are no isolated, weakly fractal geometric crosses
that could reasonably be tracked numerically.
Consequently, we cannot make meaningful statements
about a globally recrossing-free \ac{DS}.


\subsection{Slow driving frequency}
\label{sec:ps/slow}

\begin{figure*}
    \includegraphics[width=\textwidth]{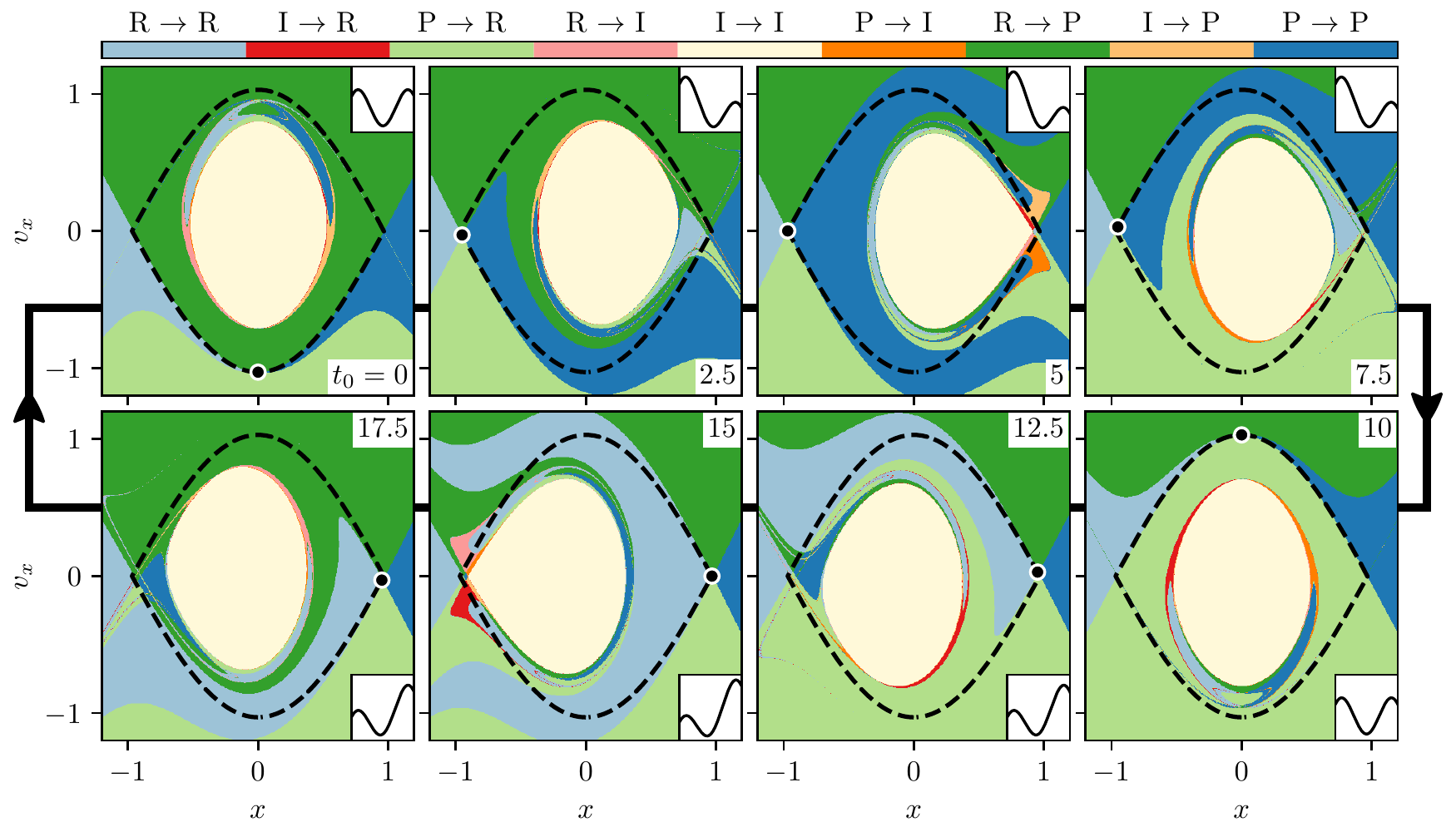}
    \caption{%
        Reactive regions for potential~\eqref{eq:materials/model}
        with $\omega = \pi / 10$ and $\nmc = 8$
        as a function of time $t_0$ (\cf\ inset labels).
        The color (or shaded) legend at the top
        indicates whether a particle starts or ends
        in the reactant (R), intermediate (I), or product (P) state.
        The position of the point on the \acs{NHIM}
        associated with the primary geometric cross (black dot)
        is tracked across a full period
        (\acs{TS} trajectory, black dashed line).
        The insets in the upper right corner of the top row panels and
        the lower right corner of the bottom row panels illustrate
        the potential at the corresponding initial time $t_0$.}
    \label{fig:reactive_regions}
\end{figure*}

The previous sections have shown
the range of complexity the model system~\eqref{eq:materials/model} can exhibit.
We now need to move
from the aesthetically pleasing structures of Fig.~\ref{fig:fractal_ld}
to a more rigorous identification of the globally recrossing-free \ac{DS}.
To do so, we switch to a lower driving frequency $\omega = \pi / 10$
which is simpler to analyze but still exhibits nontrivial behavior.
Additionally, we employ the concept of reactive regions
as described in Sec.~\ref{sec:materials/geometry}
instead of the \ac{LD}.
The partitioning of the phase space into nine distinct regions
allows to make quantitative assessments more easily.

Application of this analysis to $V(x, t)$ with $\omega = \pi / 10$
leads to the time-dependent regions
shown in Fig.~\ref{fig:reactive_regions}.
Although fractal-like structures still remain,
they are less pronounced
and mostly concentrated around whichever saddle happens to be the
lower saddle at a given instance.
The higher saddle, on the other hand, is accompanied by a
clearly visible geometric cross,
where the four regions known from the one-saddle case
(no intermediate states) meet.
The regions are also arranged in the same way:
\RR{R}{P} on top, \RR{R}{R} to the left,
\RR{P}{P} to the right, and \RR{P}{R} below.
In the following,
we refer to this geometric cross as the \emph{primary geometric cross}.


\section{The geometric cross}
\label{sec:cross}

We will now analyze the primary geometric cross
and its associated \ac{TS} trajectory in more detail.
Its time-dependent position can be tracked precisely over a full period
using the algorithm described
in Sec.~\ref{sec:materials/geometry} and Appendix~\ref{sec:bcm}.
The result is marked as a series of black dots
in Fig.~\ref{fig:reactive_regions}.
The corresponding trajectory connecting those dots
is indicated as a black dashed line.


\subsection{Global transition state trajectory}
\label{sec:cross/ds}

The primary geometric cross associated with the
instantaneously higher saddle
in Fig.~\ref{fig:reactive_regions}
remains on the barrier nearly
as long as the barrier remains dominant.
However, when the barriers' heights approach each other,
the geometric cross quickly moves from one barrier to the other
in the following way.
The geometric cross begins to rapidly accelerate towards the middle ($x = 0$).
It crosses the local potential minimum
exactly when both saddle points are level (\eg\ $t_0 = 10$)
and continues in the same direction
until it is located near the now higher saddle (\eg\ $t_0 = 15$).
The reverse happens in the following half period,
thereby forming a closed trajectory
with the same period $T_1 = 20$ as the potential $V(x, t)$
(dashed line in Fig.~\ref{fig:reactive_regions}).

\begin{figure}
    \includegraphics[width=\figurewide]{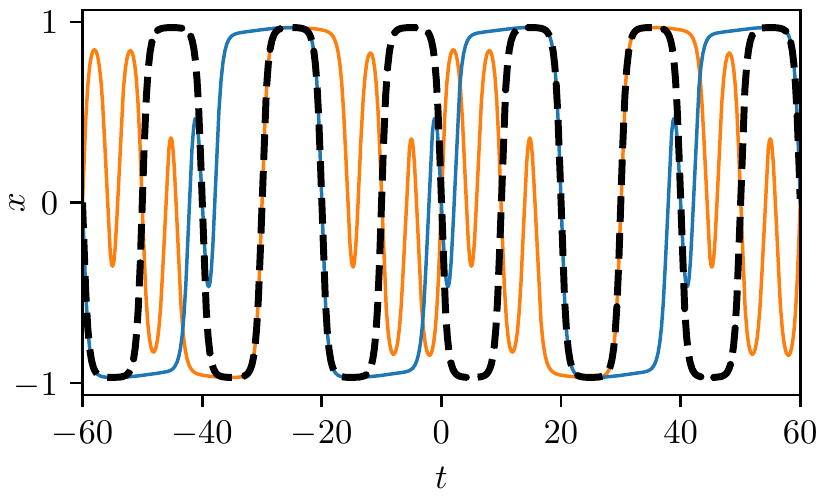}
    \caption{%
        Examples of typical periodic trajectories $x(t)$
        with periods $T_2 = 40$ (darker gray or blue)
        and $T_3 = 60$ (lighter gray or orange)
        for potential~\eqref{eq:materials/model} with $\omega = \pi / 10$.
        The primary \acs{TS} trajectory
        from Fig.~\ref{fig:reactive_regions}
        with period $T_1 = 20$ (black dashed line)
        is shown for comparison.}
    \label{fig:periodic}
\end{figure}

This primary geometric cross
marks the position of a particle on an unstable periodic trajectory
trapped in the interaction region.
The particle on this trajectory
oscillates between the saddles with a period of $T_1 = 20$
so that it is always located near the higher saddle.
Many other unstable periodic trajectories associated
with geometric crosses---\ie\ hyperbolic fixed points---in phase space
have also been found for particles in the interaction region.
Two such trajectories are shown in Fig.~\ref{fig:periodic}.
They are typical of an increased degree of structure
relative to the primary \ac{TS} trajectory.
All such trajectories together
form the system's disjoint, fractal-like time-dependent \ac{NHIM}.
In contrast to the primary \ac{TS} trajectory, however,
all other trajectories have periods larger than $T_1$.
The only exceptions to this observation
are the local \ac{TS} trajectories in the vicinities of the saddle maxima,
which are also part of the global \ac{NHIM}.
These trajectories have the same period $T_1$ as the potential by construction.

\begin{figure}
    \includegraphics[width=\figurewide]{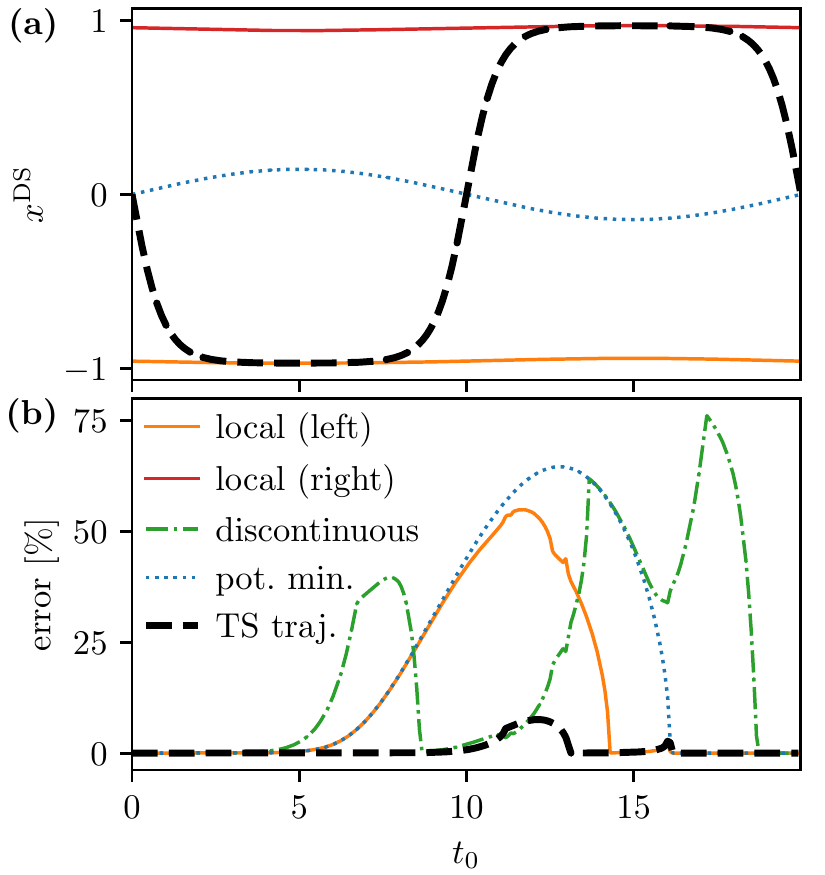}
    \caption{%
        (a)~Position \xDS\ and
        (b)~percentage of trajectories with recrossings or classification errors
        as a function of time $t_0$ for various choices of their assignment
        as reactive or nonreactive.
        The assignment is performed according to the crossing of a specified
        \acs{DS} associated with either
        the local (solid) or global (thick dashed) \acs{TS} trajectories,
        the instantaneous potential minimum between the saddles (dotted),
        or the discontinuous \acs{TS} trajectory jumping
        between the local ones (dash-dotted).
        For every $t_0$, an ensemble of \num{e6} particles with
        uniformly distributed velocities $\num{1.7} \le v_x(t_0) \le \num{2.1}$
        is initialized at $x(t_0) = \num{-3}$
        and propagated for $\Delta t = 80$ time units.
        The attached \acs{DS} is parallel to $v_x$ for all times.
        The right local \acs{DS} is recrossing-free by construction
        since particles are started solely to the left of the saddles
        and is therefore not shown in (b).
        This specific choice for the ensemble
        is also the reason for the graph's asymmetry.}
    \label{fig:ds_comparison}
\end{figure}

The complex dynamics that arises from the moving barriers is also affected
by the degree to which a given trajectory is decoupled from the barriers
as it traverses the well between them.
Although the energy of the minimum stays roughly constant, as indicated
in Fig.~\ref{fig:reactive_regions}, its position moves back and forth
between the barriers as shown in Fig.~\ref{fig:ds_comparison}(a)
so that it is always closer to the lower barrier.
This is a manifestation of the Hammond postulate~\cite{hammond55}
but now applied to the negative potential.
When the first (left) barrier dominates the dynamics,
a locally nonrecrossing \ac{DS} associated with it identifies
the trajectories with sufficient energy to cross it, and those continue
unabated across the well and the second barrier.
As a consequence, a \ac{DS} located at the well identifies
the reactive trajectories equally well (or badly) in this regime.
This is indicated in Fig.~\ref{fig:ds_comparison}(b)
by good identification when the activated particle continue past the second
barrier
or disagreement when misidentification
of trajectories begin to be reflected across both.
When the second (right) barrier dominates the dynamics,
identifying reactive trajectories at a \ac{DS} at the distant well
allows the evolving trajectories to be reflected by the barrier leading to
recrossings.
Thus we find that the reaction dynamics in between the
barriers---\eg\ at the well---does not go through a single
identifiable doorway.
In turn, this points to the need
for describing the dynamics---even in a local sense---through
a geometric picture that spans the two barriers,
\eg\ the global NHIM.

Meanwhile, since the \ac{NHIM} now consists of more than one trajectory,
it is not necessarily obvious which of these is most suited
for attaching a \emph{global} \ac{DS}.
We can, however, set conditions
the global \ac{TS} trajectory should fulfill.
First, for symmetry reasons,
the trajectory should have the same period $T_1 = 20$ as the potential.
Second, the global \ac{TS} trajectory should approach
the higher saddle's local \ac{TS} trajectory
in the (quasi-)static limit $\omega \to 0$ (\cf\ Sec.~\ref{sec:ps}).
The only trajectory found that matches both criteria
is the primary \ac{TS} trajectory introduced previously,
see Fig.~\ref{fig:ds_comparison}(a).
The large featureless regions surrounding the trajectory in phase space
additionally suggest that it affects
a significant fraction of the system's dynamics.
We will therefore refer to this trajectory as the global \ac{TS} trajectory.


\subsection{Comparison of dividing surfaces}
\label{sec:cross/quality}

The next task is to determine the
degree to which the global \ac{TS} trajectory
gives rise to a recrossing-free \ac{DS}.
Numerically, this can be tested by attaching
a \ac{DS} to it as specified in the caption of Fig.~\ref{fig:ds_comparison},
propagating an ensemble initialized near it,
and recording the number and direction of \ac{DS} crossings (or not)
that transpire thereafter in the propagated trajectories.
For simplicity, we consider only
the most challenging cases in which
the ensemble's initial energies are chosen to be
between the saddles' minimum and maximum heights.
The usual error in the \ac{DS} is signaled by the
existence of more than one crossing for the trajectories,
and the fraction of such recrossings is used below
as a measure for the \ac{DS}'s quality.

For simplicity, we limit ourselves to \acp{DS} defined by
$x = \xDS(t)$, \ie, parallel to $v_x$.
The results of this analysis are shown in Fig.~\ref{fig:ds_comparison}(b).
As can be seen, the global \ac{DS}
associated with the time-dependent geometric cross
features error rates that are significantly reduced
compared to the local \ac{DS} fixed at the left barrier.
One possible \ac{DS} could be constructed by placing it at
the instantaneous potential minimum---shown as the dotted curve
in Fig.~\ref{fig:ds_comparison}(a).
It would be expected to be ineffective given that rates are usually
determined by rate-limiting barriers, not valleys, in between
reactants and products.
Indeed, the recrossing errors found for this
\ac{DS}, shown in  Fig.~\ref{fig:ds_comparison}(b),
were high and even worse than those from the use of the \ac{DS}
fixed at the left barrier.
But the highest error rate comes from the naive attempt to
treat the \ac{DS} associated with the instantaneously higher saddle point
as the global one.
In this case, errors can arise from events beyond the recrossing of the \ac{DS}.
That is, there now exists the possibility that
the discontinuous instantaneous \ac{DS} can jump over the trajectory.
It is the combination of recrossing and classification errors that leads
to the jagged and large deviations in the \si{\percent} error
seen for the naive discontinuous \ac{DS}.
As discussed in Appendix~\ref{sec:discont_classerr},
this can lead to
\emph{misclassification} of the reactivity of the trajectory.

Finally, we consider a local \ac{DS} fixed at the right barrier.
This choice would lead to no recrossing or classification errors
for trajectories moving in the forward direction (from reactants to products).
However, it would fail badly for trajectories moving in the backward direction
by symmetry with our finding for the forward trajectories crossing the
\ac{DS} at the left barrier.
Thus, the use of the global \ac{DS} associated with the \ac{TS} trajectory
best captures the time-dependent geometry of the reaction.

\begin{figure}
    \includegraphics[width=\figurewide]{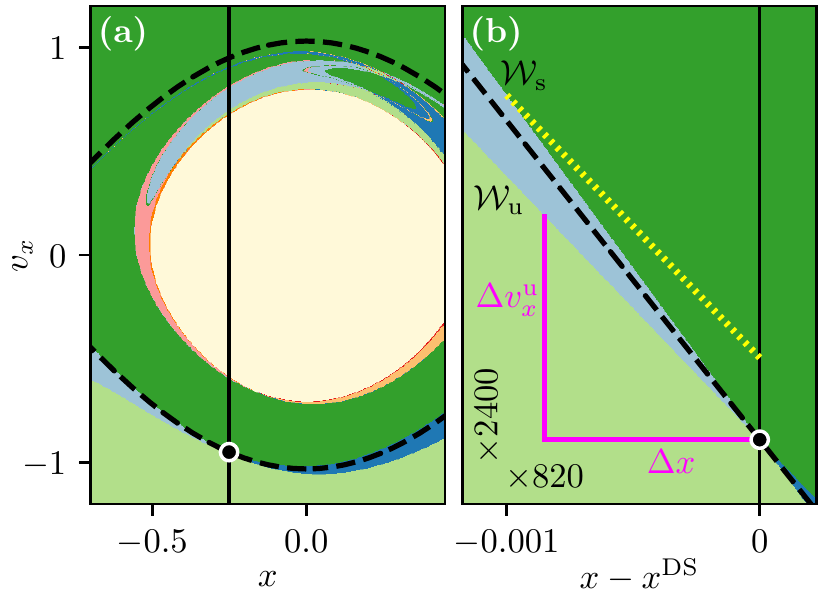}
    \caption{%
        Reactive regions analogous to Fig.~\ref{fig:reactive_regions}
        at $t_0 = \num{0.25}$
        with the addition of the \acs{DS} (solid black line).
        (a)~The simplest choice of a \acs{DS} parallel to $v_x$
        leads inevitably to recrossings as indicated.
        (b)~The immediate region of the \acs{TS} trajectory
        [shown as a dot in panel (a)],
        is enlarged, as indicated, to reveal
        the geometry of the ensemble (densely dotted)
        used in the rate calculations of Sec.~\ref{sec:rates}.
        The ensemble is sampled equidistantly
        on a line parallel to the unstable manifold \Wu\
        at distance $x - \xDS = \num{-1e-3}$
        from the \acs{TS} trajectory at $\xDS(t_0)$.
        The manifolds \Wu\ and \Ws\ are given
        by the boundaries between reactive and nonreactive regions.
        The differential of \Wu\ is indicated by
        the sides $\Delta x$ and $\Delta v_x^\mathrm{u}$
        of the slope.}
    \label{fig:recross_regions}
\end{figure}

Although this \ac{TS}-trajectory \ac{DS}
is much better than any alternatives considered so far,
it still exhibits an amount of recrossings
that cannot be explained by numerical imprecision alone.
Instead,
recrossings are caused by the fractal-like phase space structure of the system:
Figure~\ref{fig:recross_regions}(a), for example,
shows the phase space structure at $t_0 = \num{0.25}$.
We can see a relatively large patch of nonreactive reactants
(labeled \RR{R}{R}, light blue)
to the right of the \ac{DS}.
Particles in this patch leave the interaction region to the reactant (left) side
forwards and backwards in time.
Consequently, they need to cross the \ac{DS} at least twice,
which counts as a recrossing.
An analogous argument can be applied
to \RR{P}{P} regions to the left of the \ac{DS}.
The fractal-like phase space structure thus leads to
numerous problematic patches of vastly different sizes,
hence the two distinct peaks in Fig.~\ref{fig:ds_comparison}(b).

A totally recrossing-free \ac{DS}, by contrast,
would necessarily have to divide the phase space
such that \RR{R}{R} regions are always on the reactant side,
and \RR{P}{P} regions always on the product side.
This is not possible with a planar \ac{DS} of any orientation
due to the system's fractal-like nature.
A globally recrossing-free \ac{DS}---if it exists---would have to
curve as time passes
because the entirety of the phase space between the saddle points
has a clockwise rotating structure
(including periodic trajectories on the \ac{NHIM},
\cf\ Figs.~\ref{fig:reactive_regions} and~\ref{fig:periodic}).
The periodicity of this system
would then result in a fractal, spiral-like \ac{DS}.
Thus the next step in generalizing this theory would require
the identification of a non-planar \ac{DS} anchored at the
\ac{TS} trajectory which we leave as a challenge to future work.


\section{Reaction rate constants}
\label{sec:rates}

\begin{figure}
    \includegraphics[width=\figurewide]{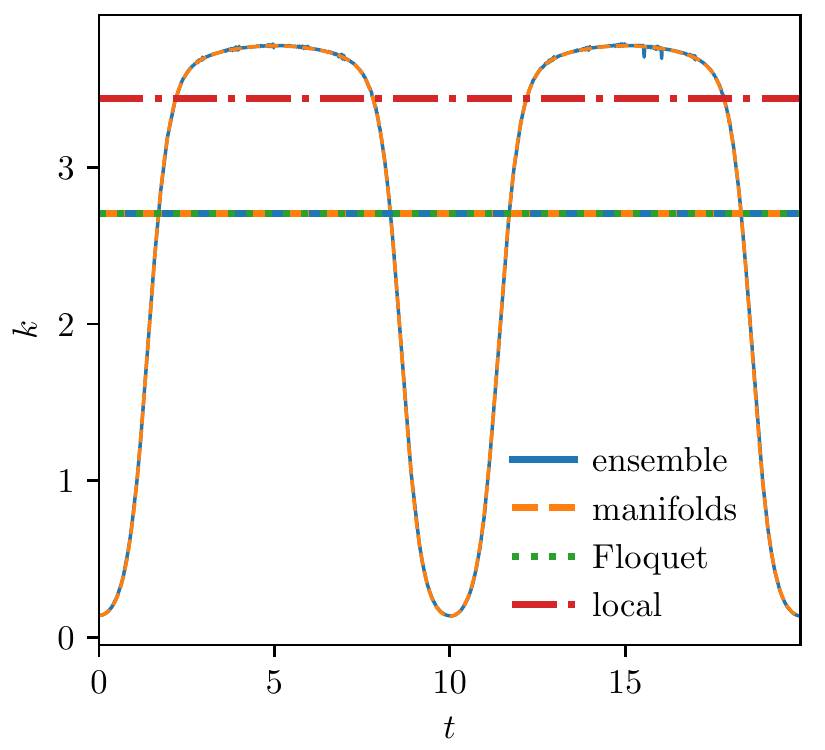}
    \caption{%
        Various instantaneous reaction rates $k$ parameterized by time $t$
        and associated with the global \acs{TS} trajectory
        of the potential defined in Eq.~\eqref{eq:materials/model}
        with $\omega = \pi / 10$, and driving period $T_1 = \num{20}$.
        The ensemble rate $\kEnse(t)$ (blue solid)
        is obtained by propagation of 20 ensembles of \num{e5} particles each
        as described in Appendix~\ref{sec:ratemethods/ensemble}.
        The manifold rate $\kMani(t)$ (orange dashed)
        is obtained using Eq.~\eqref{eq:ratemethods/k_m}.
        The Floquet rate constant \kFloq\ (green dotted)
        is obtained from Eq.~\eqref{eq:ratemethods/k_f}.
        The mean rates, \kEnseAvrg\ and \kManiAvrg\
        (thick horizontal lines)
        are averaged over the period
        as defined in Eq.~\eqref{eq:materials/k_e_avrg}.
        For comparison, the Floquet rate constant \kFloqLoc\ (red dash-dotted)
        of a single barrier's local \acs{TS} trajectory is also shown.}
    \label{fig:rate_constant}
\end{figure}

We can now calculate decay rate constants $k$
for the activated complex
using the global \ac{TS}-trajectory \ac{DS} defined in Sec.~\ref{sec:cross}.
The patches leading to recrossing
(\cf\ Sec.~\ref{sec:cross/quality})
are unproblematic in this case
because the ensemble was selected close to the \ac{NHIM} and
thereby necessarily far from them,
as can be seen in Fig.~\ref{fig:recross_regions}.
The few recrossings that do still occur
are artifacts from the numerical error in the propagation,
and are sufficiently small in number that their effect is smaller than the
numerical precision of the calculation.

In the following,
we consider the three different possible approaches
defined in Sec.~\ref{sec:materials/rates}
in order to demonstrate their equivalence in multi-saddle systems.
An example for the initial reactant ensemble and manifold geometry
can be found in Fig.~\ref{fig:recross_regions}(b).

While the application of the \ac{LMA} and the Floquet method
to our model system are straightforward,
applying the ensemble method poses a challenge:
A finite ensemble of reactants---\eg\ of size \num{e5} as implemented here---%
will mostly react within a short time%
---\eg\ \num{2} to \num{5} units of time in the case shown here---%
compared to the period of driving $T_1 = 20$.
Resolving the whole period with a single ensemble
would therefore require an exponentially growing ensemble size.
This would not be numerically feasible.
Instead, multiple ensembles have been started at
times $t_j$ incremented at equal intervals
$\Delta t$, and set to $1$ in the current case.
An instantaneous rate $\kEnse(t; t_j)$ can be obtained for each ensemble $j$.
The instantaneous rate for the whole period
$\kEnse(t)$---independent of $t_j$---can then be recovered
by concatenating the segments of each $\kEnse(t; t_j)$ for $t$
from $t_j$ to $t_{j + 1}$.
Compared to the first option,
this approach scales linearly with the system's period instead of exponentially
leading to vastly decreased computing times and increased numerical stability.

The results are shown in Fig.~\ref{fig:rate_constant}.
Since $\kEnse(t)$ and $\kMani(t)$ are not constant in time,
we additionally show the average
\begin{equation}
    \label{eq:materials/k_e_avrg}
    \kAvrg = \frac{1}{T_1} \int_0^{T_1} \dd{t} k(t)
\end{equation}
over one period $T_1$ of the \ac{TS} trajectory.

Figure~\ref{fig:rate_constant} shows a large variation
in the instantaneous reaction rate
in the interval $\num{0.13} \lesssim k(t) \lesssim \num{3.85}$.
Two features are distinctive.
First, the rate $k(t)$ shows mostly flat plateaus
during the time the \ac{TS} trajectory is located near a saddle.
This can be explained using Fig.~\ref{fig:reactive_regions}:
Although the phase space structure as a whole is undergoing
significant changes during, \eg, $\num{2.5} \le t \le \num{7.5}$,
the local vicinity around the \ac{TS} trajectory stays almost unaffected.
Second, there are deep dips in $k(t)$
while the \ac{TS} trajectory moves between the saddles
(as seen at around $t \in \qty{10 j \mid j \in \mathbb{Z}}$).
As can be seen in Figs.~\ref{fig:reactive_regions} and~\ref{fig:recross_regions},
these times are characterized by much more shallow geometric crosses
with a low difference in the slopes of the stable and unstable manifolds.
This effect is particularly apparent in Fig.~\ref{fig:recross_regions}(b).

The same observations can also be interpreted another way.
By comparing Fig.~\ref{fig:rate_constant} to Fig.~\ref{fig:ds_comparison}(a),
we can see a clear correlation
between the velocity of the \ac{TS} trajectory and the instantaneous rate $k(t)$:
the faster the \ac{TS} trajectory moves, the lower the rate drops.

\begin{table}
    \caption{%
        Values of the reaction rate constants
        discussed in Fig.~\ref{fig:rate_constant}.
        The averaged ensemble and \acs{LMA} rate constants
        \kEnseAvrg\ and \kManiAvrg\
        match the Floquet rate constants
        to within less than \SI{0.1}{\percent}.
        The local (single barrier) Floquet rate constant, however,
        differs by \SI[retain-explicit-plus]{+27}{\percent}.}
    \label{tab:rate_constant}
    \begin{tabular}
        {l l S[round-mode=figures, round-precision=5, table-format=2.4]}
        \toprule
        Method                          & Symbol     & {Value}            \\
        \midrule
        ensemble propagation            & \kEnseAvrg & 2.7055             \\
        manifold geometry (\acs{LMA})   & \kManiAvrg & 2.7062045871555145 \\
        Floquet stability analysis      & \kFloq     & 2.7036088391591266 \\
        \midrule
        local saddle (Floquet analysis) & \kFloqLoc  & 3.43837170284471   \\
        \bottomrule
    \end{tabular}
\end{table}

As can be seen
in Fig.~\ref{fig:rate_constant} and Table~\ref{tab:rate_constant},
\kEnseAvrg, \kManiAvrg, and \kFloq\ are in excellent agreement,
which illustrates the equivalence of all three methods.
The local Floquet rate constant \kFloqLoc\ of a single saddle,
on the other hand, differs significantly from \kFloq,
even though both saddles are identical.
While the local rate constant can thus be used
as an upper limit for the overall rate constant,
there is no straightforward way to derive a global rate constant from it.
Thus global methods employing the full \ac{TS} trajectory
are necessary if accurate rates for multi-saddle systems are desired.
All three methods in this section satisfy this requirement,
and are consequently in agreement.


\section{Conclusion and outlook}
\label{sec:conclusion}

In this paper, we have
fully characterized the reaction geometry and
determined the associated decay rates
in oscillatory (or time-dependent) two-saddle systems
in the framework of \ac{TST}.

The first set of central results of this work
lies in revealing the phase space structure of
the two-saddle model system.
While the structure of stable and unstable manifolds
is straightforward for (quasi-)static ($\omega \to 0$)
or very fast oscillating ($\omega \to \infty$) systems,
intermediate frequencies lead to a fractal-like phase space.
In this case,
the existence of a completely recrossing-free \ac{DS} is questionable.
For lower oscillation frequencies, however,
an isolated geometric cross
with negligible substructure---%
referred to as the \emph{primary} geometric cross---emerges.
This structure is part of the \ac{NHIM}
and can now be referred to as a \emph{global} \ac{TS} trajectory
in contrast to the \emph{local} \ac{TS} trajectories associated
with the respective single barriers.
The global \ac{TS} trajectory
oscillates between the two local \ac{TS} trajectories
with the same frequency as the potential
and can be used to attach a mostly recrossing-free \ac{DS}.

The second set of central results of this work
involves the determination of the decay rate constants of
the oscillatory (or time-dependent) two-saddle model system.
Using the \ac{DS} acquired in the first part,
we can propagate ensembles of particles,
record a time-dependent reactant population,
and finally derive an instantaneous reaction rate parameterized by time
according to Ref.~\cite{hern19e}.
Alternatively, the same result can be achieved
purely by analyzing the time-dependent phase space geometry.
For comparison,
a rate constant can be obtained from the global \ac{TS} trajectory
by means of Floquet stability analysis.
This method is in excellent agreement
with the average of each instantaneous rate.

While these results mark an important step
in the treatment of time-dependent multi-saddle systems,
many questions still remain unanswered:
First, we restricted ourselves to two-saddle systems with one \ac{DoF}.
To be applicable to real-world systems, however,
the methods presented here
will have to be generalized to at least more \acp{DoF}
because few chemical reactions can be treated exactly, or nearly so,
when reduced to just one coordinate.
Second, it will be important
to investigate the influence of minor manifold crossings
on the rate constant.
This is particularly necessary for
cases of time-dependent barriers found here in which
there is no longer an equivalent to the primary geometric cross.
This is even more challenging when the alternation between barriers
is driven at high frequencies (\cf\ Fig.~\ref{fig:fractal_ld}).

The applicability of our results to real-world systems also remains to
be demonstrated.
It remains unclear whether it is possible
to treat systems without a primary geometric cross.
That chemical reactions can be represented with potentials
exhibiting the challenges discussed here, is illustrated by
the isomerization of ketene via formylmethylene and oxirene
which has been modeled via a four-saddle
potential~\cite{gezelter1995, hern13c, wiggins14b, hern14e, hern16d}.
The isomerization reaction of triangular KCN
via a metastable linear K-CN configuration
can similarly be described by a two-saddle system~\cite{borondo13a, borondo18a}.
Thus the analysis of time-dependent driven potentials resolved here,
when applied to these and other chemical reactions,
should provide new predictive rates for driven chemical reactions of interest.


\begin{acknowledgments}
    Useful discussions with Robin Bardakcioglu are gratefully acknowledged.
    The German portion of this collaborative work was partially supported
    by the Deutsche Forschungsgemeinschaft (DFG) through Grant No.~MA1639/14-1.
    The US portion was partially supported
    by the National Science Foundation (NSF) through Grant No.~CHE 1700749.
    MF is grateful for support from
    the Landesgraduiertenförderung of the Land Baden-Württemberg.
    This collaboration has also benefited from support
    by the European Union's Horizon 2020 Research and Innovation Program
    under the Marie Skłodowska-Curie Grant Agreement No.~734557.
\end{acknowledgments}


\appendix

\section{Tracking geometric crosses}
\label{sec:bcm}

\begin{figure}
    \includegraphics[width=\figurewide]{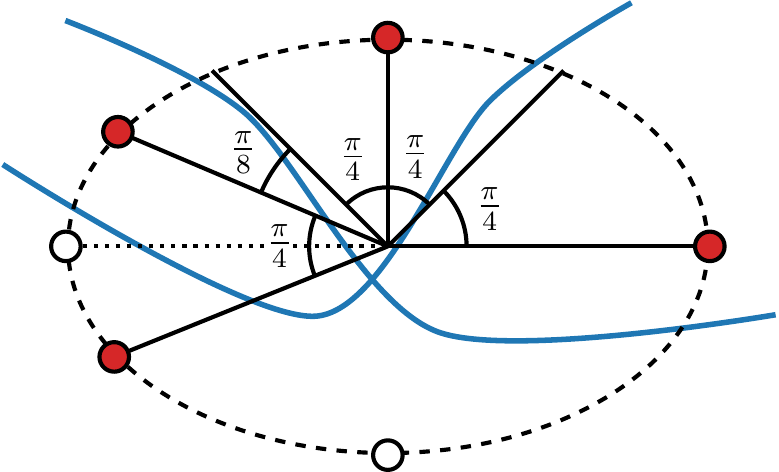}
    \caption{%
        Sketch of the algorithm for finding the inputs to the \acs{BCM}:
        four arbitrary initial phase space coordinates
        representative of the four regions associated with the \acs{NHIM}.
        Instead of naively choosing coordinates
        on two right-angled axes (empty circles),
        an ellipse enclosing the geometric cross is constructed.
        The first point (filled circle on right hand side)
        is selected arbitrarily at coordinates with angle $\alpha = 0$.
        Subsequent coordinates are found by incrementing $\alpha$
        in steps of $\Delta \alpha = \pi / 4$ (solid lines),
        and confirming that said point satisfies the condition of a new region.
        If a region is skipped---\eg\ as shown in the dotted line---the
        increment $\Delta \alpha$ is temporarily reduced
        until a point in the region in between is found.}
    \label{fig:bcm_init_quad}
\end{figure}

To be able to track geometric crosses in phase space reliably,
the \acf{BCM} (\cf\ Sec.~\ref{sec:materials/geometry}) needs to be initialized
with four points in four different reactive regions
without human intervention.
Since the geometric crosses of interest can be quite distorted
while moving between saddles,
the \ac{BCM} cannot be used
precisely in the way described in Ref.~\cite{hern18g}.
Originally, the initial quadrangle was defined
by guessing the geometric cross's position and choosing two coordinates each
on a horizontal and a vertical line through it.
In our case, though, we first define an ellipse enclosing the geometric cross,
centered on this guess.
The initial corners of the quadrangle are then selected on the ellipse
according to the algorithm described in Fig.~\ref{fig:bcm_init_quad}.

Interestingly, even a given geometric cross
is not entirely free of substructures.
Instead, it features a fractal-like set of crossing manifolds
in its close proximity.
Since these additional structures are extremely thin, however,
they do not hinder the \ac{BCM}
from finding the desired geometric cross coordinates
up to the desired precision
which maybe be even smaller than the width of substructures.


\section{Decay rates methods}
\label{sec:ratemethods}


\subsection{Ensemble method}
\label{sec:ratemethods/ensemble}

The conceptually simplest method for calculating decay rates \kEnse\
is by means of propagation of an ensemble.
In analogy to Ref.~\cite{hern19e}
we first identify a line segment parallel to the unstable manifold
that satisfies the property:
it lies on the reactant side between the stable manifold and the \ac{DS}
at a distance that is small enough to allow for linear response
and large enough to suppress numerical instability.
At $t = t_0$, an ensemble of particles is placed on this line
and propagated in time
to yield a time-dependent reactant population $\Nr(t; t_0)$.

In the second major step, one can obtain a reaction rate constant
\kEnse\ by fitting an exponential decay
$\Nr(t; t_0) \propto \exp[-\kEnse (t - t_0)]$
to the reactant population~\cite{
    pech81, chan78, mill83c, rmp90, connors90, upadhyay06}.
This, however, is not possible in all systems---and, in particular,
in the two-saddle system being investigated here---because
the decay in $\Nr(t; t_0)$ can be nonexponential.
Instead, we use the more general approach described in Ref.~\cite{hern19e},
which involves examining the instantaneous decays
\begin{equation}
    \label{eq:ratemethods/k_e}
    \kEnse(t; t_0) = -\frac{1}{\Nr(t; t_0)} \dv{t} \Nr(t; t_0)
    \MathPeriod
\end{equation}
An analogous definition has been used in Ref.~\cite{Lehmann00a}
for escape rates over a potential barrier.


\subsection{Local manifold analysis}
\label{sec:ratemethods/manifold}

In general, the approach described above is computationally expensive
since a lot of particles have to be propagated.
This led us to develop a second method, called the \ac{LMA},
for obtaining instantaneous reaction rates
purely from the geometry of the stable and unstable manifolds in phase space.
If the slopes of the stable and unstable manifolds \Ws\ and \Wu\ at time $t$
are given by $\Delta v_x^\mathrm{s}(t) / \Delta x$
and $\Delta v_x^\mathrm{u}(t) / \Delta x$
then the instantaneous rate $\kMani(t)$ can be written as
their difference~\cite{hern19e},
\begin{equation}
    \label{eq:ratemethods/k_m}
    \kMani(t)
    = \frac{\Delta v_x^\mathrm{u}(t) - \Delta v_x^\mathrm{s}(t)}{\Delta x}
    \MathPeriod
\end{equation}
This allows for the calculation of $\kMani(t)$
independently at different times $t$,
making it easy to compute in parallel.


\subsection{Floquet method}
\label{sec:ratemethods/floquet}

Last, average decay rates \kFloq\ can also be obtained directly using
a Floquet stability analysis~\cite{hern14f, hern19e}
that was seen earlier to lead to accurate rates in
reactions with a time-dependent barrier.
While this method is the computationally cheapest,
it cannot yield instantaneous rates.

To obtain the time-independent rate constant \kFloq\ for
a given trajectory on the \ac{NHIM}, we can
linearize the equations of motion using the Jacobian
\begin{equation}
    J(t) = \mqty(0 & 1 \\ -\dv*[2]{V(x, t)}{x} & 0)
    \MathPeriod
\end{equation}
By integrating the differential equation
\begin{equation}
    \dv{\sigma(t)}{t} = J(t) \sigma(t)
    \qq{with} \sigma(0) = \mathbb{1}
    \MathComma
\end{equation}
one obtains the system's fundamental matrix $\sigma(t)$.
When considering trajectories with period $T$,
$M = \sigma(T)$ is called the monodromy matrix.
Its eigenvalues $m_\mathrm{u}$ and $m_\mathrm{s}$, termed Floquet multipliers,
can be used to determine the Floquet rate constant
\begin{equation}
    \label{eq:ratemethods/k_f}
    \kFloq = \frac{1}{T} \qty(\ln\abs{m_\mathrm{u}} - \ln\abs{m_\mathrm{s}})
    \MathPeriod
\end{equation}


\section{Possible errors
    for the discontinuous DS in Fig.~\ref{fig:ds_comparison}(b)}
\label{sec:discont_classerr}

\begin{figure}
    \includegraphics[width=\figurewide]{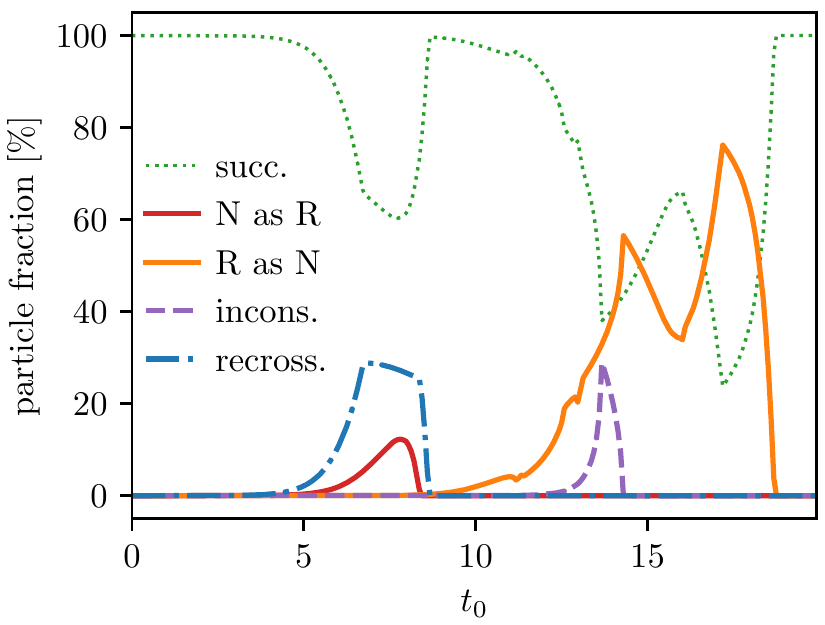}
    \caption{%
        Fraction of trajectories
        with (see text) and without (labeled \enquote{succ.}) errors
        as a function of time $t_0$
        for the discontinuous \ac{DS} from Fig~\ref{fig:ds_comparison}.}
    \label{fig:discont_classerr}
\end{figure}

There are multiple ways in which a discontinuous \ac{DS}
can show errors for a trajectory:

\emph{(i)}
    A nonreactive trajectory is classified as reactive
    (labeled \enquote{N~as~R} in Fig.~\ref{fig:discont_classerr}).
    This can happen when a reactant enters the central region
    between the saddles while the \ac{DS} is located near the left saddle.
    If this happens shortly before the \ac{DS} jumps to the right,
    the particle can get reflected at the right barrier
    and leave the central region on the reactant side.

\emph{(ii)}
    A reactive trajectory is classified as nonreactive
    (labeled \enquote{R~as~N} in Fig.~\ref{fig:discont_classerr}).
    This can happen when a reactant enters the central region
    while the \ac{DS} is located near the right saddle.
    The \ac{DS} can then jump discontinuously over the particle
    to the left saddle---which is not counted as a crossing---before
    the particle leaves the central region to the right as a product.

\emph{(iii)}
    The classification is thus inconsistent
    (labeled \enquote{incons.} in Fig.~\ref{fig:discont_classerr}).
    This can happen, \eg, when a reactant enters the central region
    while the \ac{DS} is located near the right saddle,
    and leaves it again to the reactant side
    while the \ac{DS} is near the left saddle.
    As a result, a backward reaction is recorded
    even though the particle started as a reactant.
    Similarly, it is possible to detect two forward reactions
    over the same \ac{DS} without an intermediate backward reaction.
    It is unclear which reaction is to be counted as the \emph{real} one.

\emph{(iv)}
    The trajectory crosses the \ac{DS} multiple times,
    \ie, it exhibits recrossings
    (labeled \enquote{recross.} in Fig.~\ref{fig:discont_classerr}).
    This can happen when a particle enters and leaves the central region
    on the reactant side while the \ac{DS} is near the left saddle,
    resulting in two crossings.

The distributions of the errors that
arise from the discontinuous \ac{DS}
originating from recrossing and misclassification
are shown in Fig.~\ref{fig:discont_classerr}.
The sum of these errors gives rise to that
reported in Fig.~\ref{fig:ds_comparison} for this \ac{DS}.


\bibliographystyle{apsrev4-1-forcedoi}
\bibliography{paper-q20}

\end{document}